\pdfoutput=1
\documentclass{JFM-FLM_Au}

\usepackage[usenames,dvipsnames]{xcolor}
\def\del#1{\textcolor{Gray}{}}

\def\ad{^{\dagger}}

\newcommand{\ea}[1]{\left\langle #1 \right\rangle}
\newcommand{\mea}[1]{\overline{#1}}
\newcommand{\mU}{\overline{U}}
\newcommand{\ip}[2]{\mathcal{I}\!\left(#1, #2\right)}

\usepackage{graphicx}
\usepackage{floatrow}
\usepackage{upgreek}
\usepackage{subfigure}
\usepackage{amsmath}
\usepackage{amssymb}
\usepackage{mathrsfs}  
\usepackage{nomencl}
\usepackage{eqnarray}
\usepackage{setspace}
\usepackage[usenames,dvipsnames]{xcolor}
\usepackage{tikz}
\usetikzlibrary{shapes}
\usepackage{colortbl}
 \usepackage{hyperref}
\hypersetup{
	colorlinks=true, 
	linktoc=all,     
	linkcolor=blue,  
	citecolor=blue,
	urlcolor = blue
}

\lefttitle{Qi Wang and Tamer A. Zaki}
\righttitle{Journal of Fluid Mechanics}

\title{Mitigating adjoint chaos in wall turbulence}

\author{Qi Wang \aff{1,2}
\and Tamer A. Zaki\aff{1}}

\affiliation{\aff{1} Department of Mechanical Engineering, Johns Hopkins University, Baltimore, MD 21218, USA 
\aff{2} Department of Aerospace Engineering, San Diego State University, San Diego, CA 92182, USA}

\corresau{Tamer A. Zaki, \email{t.zaki@jhu.edu}}

\usepackage{bm}
\usepackage{amsmath}
\definecolor{DarkGreen}{RGB}{26, 161, 44}

\begin{document}
\maketitle

\begin{abstract}
Estimating past events in wall turbulence based solely on surface measurements and first principles is an ill-posed problem that is complicated by chaos.  The sensitivity of a measurement to the earlier flow state is described by the adjoint Navier-Stokes equations, which are solved in reverse time starting from the measurement kernel at the sensing position and time.  The resulting adjoint field is the spatio-temporal domain of dependence (DOD) of the sensor, which is a dual to the concept of the domain of influence (DOI) of an actuator in the linearized forward equations.  In channel turbulence, the energy of each adjoint realization grows exponentially in backward time according to the Lyapunov exponent, even though the energy of the ensemble average should decay.  We introduce a linear eddy-viscosity closure model in the ensemble-averaged adjoint equations, and directly compute the mean DOD and compare our prediction to the ensemble average. Furthermore, we demonstrate that the DOD of a wall-stress measurement and the DOI resulting from a wall-stress perturbation exhibit respective universal behaviors across Reynolds numbers. However, their spatio-temporal structures differ qualitatively, due to the time-asymmetry of the governing equations. The DOD field has a two-part structure: one component is associated with the Orr mechanism, characterized by rapid reorientation under mean shear, and the other is related to self-similar expanding streaky structures. These two components jointly define the sensitivity of the wall-stress measurement to past flow events.

\end{abstract}

%


\section{Introduction}

In turbulent channel flow, the wall plays an important role as a net source of vorticity injection into the domain.  When considered from an observer's perspective, the wall vorticity depends on the earlier-in-time flow state.  Previous efforts have examined the contributions to the spanwise wall-vorticity, or the streamwise surface stress,
by stretching and tiling of earlier vorticity \citep{xiang2025origin}.  In the present work, we are concerned with the utility of measurements of the wall stress to estimate the earlier flow state.  Our objective is to determine the domain of dependence (DOD) of the measurement \citep{wang2025domain}, which is defined as the earlier-in-time spatio-temporal state that can optimally alter the recorded surface stress. We develop an eddy-viscosity model to efficiently estimate the ensemble-averaged DOD in turbulent channel flow, compare it to the computationally expensive ensemble average, and demonstrate its universality and dynamics across Reynolds numbers.

The concept of a domain of dependence originates from the theory of hyperbolic partial differential equations, where the solution at a given space-time point depends only on the initial condition within a characteristic cone with boundaries defined by the characteristic speeds of the system \citep{courant1963methods}. For the Navier-Stokes equations, which are parabolic-elliptic and propagate information at infinite speed of sound, a sensitivity-based analog is more appropriate: one seeks not which points affect the measurement, but the ones that have a strong causal influence. Identifying earlier flow events causally affecting a surface measurement is central to state estimation, model-based flow control, and the physical interpretation of dynamics.
The adjoint framework is powerful for data assimilation and, closely related, the evaluation of the sensitivity of measurements to past flow events. In data assimilation, the adjoint field yields the gradient of the cost function with respect to the unknown parameters, and thus guides the minimization of the mismatch between the simulation and measurements \citep{wang_hasegawa_zaki_2019, zaki2025arfm, wang2025variational}.  In sensitivity analysis, the adjoint field is the flow state that has the greatest impact on an observable, and offers profound physical insight, particularly in linear, non-chaotic settings such as passive scalar transport in laminar flows \citep{pudykiewicz1998application,keats2007bayesian,neupauer2001adjoint, kgb6-k3zm}. 
In turbulent flows, adjoint methods have demonstrated the inherent difficulty of reconstructing the full flow field from limited measurements. When the observations are the wall stresses, only near-wall structures and the largest outer-layer motions can be reliably recovered \citep{wang_wang_zaki_2022}. 

A key challenge in turbulent flows is the exponential amplification of the adjoint field in backward time, which is the dual to the classical butterfly effect \citep{lorenz2000butterfly, deissler1987spatially}. Infinitesimally close measurements can arise due to entirely different flow states \citep{zaki2021prfaps}.  The exponential growth of each adjoint realization in backward time can render both data assimilation and sensitivity analyses unstable or infeasible. The problem is especially acute when the target quantities are long-time averages 
\citep{yuan2023adjoint,garai2021stabilization,xu2023review}.
The difficulty has its roots in the fluctuation-response relations of statistical mechanics. Kubo's theorem \citep{kubo1957statistical} states that the linear response of an equilibrium observable to a small perturbation equals an equilibrium time-correlation function of the unperturbed system. In the present context, the statistically stationary turbulent channel flow is the equilibrium ensemble, and Kubo's theorem identifies the ensemble-averaged adjoint as a two-point space-time velocity correlation of the unperturbed turbulence. Two consequences follow. First, the mean adjoint must be smooth and well-behaved even though individual realizations diverge exponentially, because the underlying equilibrium correlations of stationary turbulence are finite and regular. Second, any closure that accurately captures the dominant space-time correlations of the turbulence should yield a faithful approximation of the mean DOD without requiring direct ensemble averaging of chaotic trajectories.

To address these difficulties, stabilization strategies have been developed. Least-Squares Shadowing (LSS) replaces the ill-conditioned initial-value problem with a well-posed variational problem, yielding accurate sensitivities of long-time averages in a simple, chaotic system \citep{wang2014least, wang2014convergence, chater2017least, ni2017sensitivity}. The primal trajectory is replaced by a shadow trajectory that remains as close as possible to the original chaotic orbit while satisfying the linearized governing equations. The ill-posed backward-time initial-value problem is then converted into a coupled forward-backward boundary-value problem with a well-conditioned normal equation, eliminating the exponential divergence at the cost of a global time-window solve. LSS yields accurate sensitivities for low-dimensional chaotic systems such as the Lorenz equations and the Kuramoto-Sivashinsky equation \citep{wang2014least, chater2017least}, but its computational cost scales with both the state dimension and the time horizon, making direct application to turbulent flows prohibitive. Non-intrusive least-squares shadowing (NILSS) \citep{ni2017sensitivity} reduces this cost by restricting the shadow correction to the unstable subspace of the flow, requiring only a number of additional forward simulations equal to the number of positive Lyapunov exponents. However, the dimension of the unstable subspace grows with Reynolds number in turbulent shear flows, limiting the practical usage \citep{chandramoorthy2019feasibility}.
Adjoint-energy-based stabilization introduces artificial damping motivated by the adjoint energy budget, and offers a cheaper alternative for turbulent flows \citep{blonigan2012towards, garai2021stabilization}.

In the context of the present work, the mean DOD requires an ensemble average of the adjoint fields \citep{wang_wang_zaki_2022}, evaluated from different realizations of the forward trajectory. However, due to a power-law tail in the histogram of adjoint gradients, ensemble averaging converges very slowly \citep{eyink2004ruelle}. A potential remedy to this slow convergence is sought, and is inspired by the literature on the forward-time linearized dynamics in turbulent channel flow. 
Rather than study the forward evolution of perturbations to the full turbulent state, which are exponentially unstable, \citet{del2006linear} and \citet{cossu2009optimal} considered the mean turbulent profile and adopted an eddy-viscosity model for the analysis.  They demonstrated that, despite the damping of all eigenmodes, perturbations can exhibit significant transient energy growth that closely mirrors the organization and scaling of coherent structures observed in wall turbulence. 
\citet{vadarevu2019coherent} showed that an impulsive body force in the linearized Navierâ€“Stokes system, augmented with an eddy-viscosity model, produces vortexâ€“streak structures that exhibit geometric self-similarity and decay predictably with time.
These analyses predict an inner and outer peak in the optimal energy amplification, and a self-similar evolution of streamwise streaks. 
In addition to the growth of streaks by lift-up \citep{landahl1980note,liu2024lift}, Orr amplification is another important non-modal mechanism where perturbations that are initially tilted against the mean shear grow as they are rotated toward the direction of maximum strain.
\citet{encinar2020momentum} demonstrated that conditionally averaged burst structures in fully nonlinear direct numerical simulations (DNS) are consistent with the Orr reorientation of upstream-tilted perturbations. Their results establish a direct connection between observed turbulent structures and the non-normal amplification of linearized dynamics.
While the above studies focus on the forward-time evolution and statistical interpretation of coherent structures, the present work addresses the complementary problem of backward-time sensitivity: namely, which earlier perturbations most strongly influence a given wall-stress measurement.

The rest of the paper is structured as follows. In \S\ref{Sec:forumulation}, we present the mathematical formulation for the eddy-viscosity closure of the ensemble-averaged linear forward and adjoint equations, and provide a physical interpretation of the mean adjoint fields. In \S\ref{sec:results}, we contrast the forward and adjoint fields, and demonstrate their respective self-similar behaviors.  We compare the adjoint model to ensemble averaging of realizations, and examine the forward-time dynamics of the computed adjoint DOD to aid the interpretation.  Conclusions are provided in \S\ref{sec:conclusion}.

\section{Mathematical Formulation}
\label{Sec:forumulation}

We start from the linearized incompressible Navierâ€“Stokes (NS) equations for turbulent channel flow, and perform Reynolds-averaging to evaluate the ensemble statistics. This leads to the Reynolds-averaged linearized Navierâ€“Stokes equations (RA-LNS). With an appropriate closure model, we obtain the forward formulation and then derive the associated Reynolds-averaged adjoint Navier-Stokes (RA-ANS) system for backward-time sensitivity analysis. The reasoning and limitations of eddy-viscosity closure are discussed. The physical interpretations of these equations are then presented: the RA-LNS describes the forward-in-time domain of influence (DOI) of a disturbance, while the RA-ANS quantifies the back-in-time domain of dependence (DOD) of an observation. 
All the numerical solutions of the forward and adjoint Reynolds-averaged equations reported herein are performed in spectral space, using standard Fourier-Fourier-Chebyshev discretization of the governing equations.

\subsection{The Reynolds-averaged linearized Navier-Stokes equations}
\label{sec:RALNS}

Suppose the incompressible velocity field $\boldsymbol{U}(\boldsymbol{x},t)$ is a realization of statistically stationary turbulent channel flow.  
Variables are non-dimensionalized by the half-channel height $h^\star$, the ensemble-averaged bulk velocity $\ea{U_b^\star}$, and the fluid density $\rho^\star$, where superscript $\star$ marks dimensional quantities and $\langle \bullet \rangle$ is the ensemble average defined below. The non-dimensional Navier-Stokes equations are,

\vspace{-6pt}
\begin{subequations}
    \label{Eqn:NS}
    \begin{eqnarray}
	    {\partial_t \boldsymbol{U}} + \left(\boldsymbol{U} \cdot \boldsymbol{\nabla} \right) \boldsymbol{U} &=& - \boldsymbol{\nabla} P + \nu \nabla^2 \boldsymbol{U}, \qquad
        \boldsymbol{\nabla} \cdot \boldsymbol{U} = \boldsymbol{0},
    \end{eqnarray}
\end{subequations}
where the non-dimensional kinematic viscosity and bulk Reynolds number are related by
$Re = 1/\nu := \ea{U^{\star}_b} h^{\star} / \nu^{\star} $.
Superscript $+$ denotes normalization by viscous scales, specifically, the ensemble-averaged friction velocity $\ea{U_{\tau}^\star} := \sqrt{ \nu^\star \left({\partial \ea{U^\star}}/{\partial y^\star}\right)_{y^\star=0} }$ and lengthscale $\nu^{\star}/\ea{U_{\tau}^\star}$. The friction Reynolds number is defined as $Re_\tau:=\ea{U_\tau^\star} h^\star / \nu^\star$.

We introduce a small perturbation field $(\boldsymbol{u},p)$ superimposed on top of $(\boldsymbol{U},P)$.  This perturbation is thus governed by the linearized Navier-Stokes equations,

\vspace{-6pt}
\begin{equation}
    \label{Eqn:pLNS}
	    {\partial_t \boldsymbol{u}} + \left(\boldsymbol{U} \cdot \boldsymbol{\nabla} \right) \boldsymbol{u} + \left(\boldsymbol{u} \cdot \boldsymbol{\nabla} \right) \boldsymbol{U} = - \boldsymbol{\nabla} p + \nu \nabla^2 \boldsymbol{u},\qquad
        \boldsymbol{\nabla} \cdot \boldsymbol{u} = \boldsymbol{0},\qquad
        \boldsymbol{u}_0 = \boldsymbol{\phi},
\end{equation}
where $\boldsymbol{\phi}$ is the initial velocity perturbation.
The ensemble average of the above equations over different realizations of the turbulent field $\boldsymbol{U}$ yields,

\vspace{-6pt}
\begin{subequations}
    \label{Eqn:LNS}
    \begin{eqnarray}
	    &{\partial_t \ea{\boldsymbol{u}}} + \ea{\boldsymbol{U}} \cdot \boldsymbol{\nabla} \ea{\boldsymbol{u}} + \left(\ea{\boldsymbol{u}} \cdot \boldsymbol{\nabla} \right) \ea{\boldsymbol{U}} = -\nabla \cdot \left(\ea{\boldsymbol{U}^\prime \boldsymbol{u}^{\prime} +\boldsymbol{u}^{\prime}\boldsymbol{U}^\prime }\right) - \boldsymbol{\nabla} \ea{p} + \nu \nabla^2 \ea{\boldsymbol{u}},\qquad \\
        &\boldsymbol{\nabla} \cdot \ea{\boldsymbol{u}} = \boldsymbol{0},\qquad
        \ea{\boldsymbol{u}}_0 = \boldsymbol{\phi}. 
    \end{eqnarray}
\end{subequations}
The ensemble average $\langle \bullet \rangle$ is computed using realizations of statistically stationary turbulent base flows, and exploiting ergodicity. 
Specifically, a long trajectory of the forward flow is sampled evenly in time, with circular translation in the homogeneous $x$ and $z$ directions, to construct an ensemble of base flows $\boldsymbol{U}(\boldsymbol{x},t)$. Each member of this ensemble is then adopted as the base flow to advance the perturbation fields $\boldsymbol{u}$.
In (\ref{Eqn:LNS}), the terms $\boldsymbol{U}'\boldsymbol{u}'+\boldsymbol{u}'\boldsymbol{U}'$ are outer (dyadic) products, with components $(U_i'u_j'+u_i'U_j')$. The primed variables 
$\boldsymbol{U}^\prime \equiv \boldsymbol{U} - \ea{\boldsymbol{U}}$ is the deviation away from the mean turbulent velocity field, and $\boldsymbol{u}^\prime \equiv \boldsymbol{u} - \ea{\boldsymbol{u}}$ is the deviation from the mean perturbation field.
Both deviations satisfy the divergence-free condition. 
We note from (\ref{Eqn:pLNS}) and (\ref{Eqn:LNS}) the distinct dynamics of the evolution of an individual perturbation $\boldsymbol{u}$ and of the mean $\ea{\boldsymbol{u}}$ evaluated over different realizations of $\boldsymbol{U}$.
As shown in previous studies \citep{nikitin2018characteristics, wang_wang_zaki_2022}, the instantaneous perturbation field grows exponentially in forward time at the Lyapunov rate. 
If we take a perturbed initial condition $\boldsymbol{U}_0^{\delta} = \boldsymbol{U}_0 + \boldsymbol{u}_0$, the evolution of
this new field would gradually converge to the same statistically stationary channel-flow turbulence with the same mean, i.e.\, $\ea{\boldsymbol{U}^{\delta}}=\ea{\boldsymbol{U}}$ as time $t \rightarrow\infty$, or equivalently $\ea{\boldsymbol{U}^\prime} = 0$. As such, the ensemble-averaged difference between the perturbed and the original fields must decay at long forward times.
As for the linear perturbations, each realization $\boldsymbol{u}$ amplifies exponentially while the ensemble average $\ea{\boldsymbol{u}}(t)$ decays in time \citep{malkus1956outline, markeviciute2023improved, zhu2024resolvent}.  
This distinction between fluctuation growth and mean decay underlies the difficulty of computing the ensemble average in order to evaluate the mean of the linearized fields, which requires an exponentially growing number of samples as the time of interest is increased.

We proceed to introduce a linear eddy-viscosity model (LEVM) to close  $\ea{\boldsymbol{U}^\prime \boldsymbol{u}^{\prime} +\boldsymbol{u}^{\prime}\boldsymbol{U}^\prime }$.
In the original LEVM, the turbulent stress tensor $\ea{\boldsymbol{U}^\prime\boldsymbol{U}^\prime}$ is modelled as,

\vspace{-6pt}
\begin{equation}
\label{eq:LEVM_mean}
    -\ea{U_i^\prime U_j^\prime} = 2\nu_t \ea{S}_{ij} - \frac 23 \ea{K}\delta_{ij},
\end{equation}
where $\nu_t$ is the effective turbulent viscosity, $\ea{S}_{ij} = \frac 12\left(\partial_j \ea{U}_i + \partial_i \ea{U}_j\right)$ is the mean strain-rate tensor for the base state $\boldsymbol{U}$, and $\ea{K}=\frac 12 \ea{U_i^{\prime}U_i^{\prime}}$ is the fluctuation kinetic energy in the base state.
Here $\epsilon \ll 1$ is an order parameter characterizing the magnitude of the velocity perturbation $\boldsymbol{u}$ relative to the base-flow scale, i.e.,\,$|\boldsymbol{u}|=O(\epsilon|\boldsymbol{U}|)$; the same $\boldsymbol{u}$ appearing in equations~(\ref{Eqn:pLNS}-\ref{Eqn:LNS}) is used throughout. Introducing this perturbation into the eddy-viscosity relation yields the same statistical pattern at leading order in $\epsilon$,
\del{Adding a small perturbation of magnitude $o(\epsilon)$ to this flow field yields the same statistical pattern,}

\vspace{-6pt}
\begin{equation}
\label{eq:LEVM_perturbed}
    -\ea{(U_i^\prime + \epsilon u_i) (U_j^\prime + \epsilon u_j)} = 2\nu_t (\ea{S}_{ij} + \epsilon \ea{s}_{ij}) - \frac 23 (\ea{K} + \epsilon \ea{U_l^\prime u_l} + \epsilon^2 \ea{k})\delta_{ij}.
\end{equation}
Here $\ea{s}_{ij} = \frac 12\left(\partial_j \ea{u}_i + \partial_i \ea{u}_j\right)$ is the mean strain-rate tensor for the perturbation field and $\ea{k} = \frac 12 \ea{u_i^{\prime}u_i^{\prime}}$ is the fluctuation kinetic energy for the perturbation field $\boldsymbol{u}$.
Subtracting equations (\ref{eq:LEVM_mean}) and (\ref{eq:LEVM_perturbed}), and dividing by $\epsilon$, yields,

\vspace{-6pt}
\begin{equation}
    -\ea{U_i^\prime u_j +U_j^\prime u_i} = 2\nu_t \ea{s}_{ij}- \frac 23 \ea{U_l^\prime u_l} \delta_{ij}+\mathcal{O}(\epsilon).
\end{equation}
For clarity, we note that the repeated $l$ index implies summation (Einstein convention).
Ignoring the higher-order terms, and applying $\partial_x \ea{U} = \partial_z \ea{U} = \ea{V} = \ea{W} = 0$ in a turbulent channel flow leads to equations for the modelled mean base flow $\mU$ and mean perturbations $\mea{\boldsymbol{u}}$ and $\mea{p}$, as approximations of $\ea{U}$, $\ea{\boldsymbol{u}}$ and $\ea{p}$, 

\vspace{-6pt}
\begin{equation}
    \label{Eqn:LRANS}
	    \partial_t \mea{\boldsymbol{u}} + \mU \partial_x  \mea{\boldsymbol{u}} + \mea{v} \;d_y \mU \, \boldsymbol{e}_x = -  \boldsymbol{\nabla} \mea{p} + \nu_t \nabla^2 \mea{\boldsymbol{u}} +  \left(d_y\nu_t\right)  \left(\partial_y \mea{\boldsymbol{u}} + \nabla \mea{v}\right), \quad  \boldsymbol{\nabla} \cdot \mea{\boldsymbol{u}} = 0, \quad \mea{\boldsymbol{u}}_0 = \boldsymbol{\phi}. 
\end{equation} 
In the above expression, the pressure $\mea{p}$ includes the $\ea{U_l^\prime u_l}$ term.
This set of equations can be termed the Reynolds-averaged linearized Navier-Stokes (RA-LNS), and they have the same form as the linearized Reynolds-averaged Navier-Stokes. In other words, the ensemble average of linear perturbations to different realizations of the base flow is the same as the linear perturbation to the ensemble-averaged base flow.
The closure above implicitly assumes that the correlation term 
$\ea{\boldsymbol{U}'\boldsymbol{u}'+\boldsymbol{u}'\boldsymbol{U}'}$
depends instantaneously on the mean perturbation strain $\partial_y \mea{\boldsymbol{u}} + \nabla \mea{v}$ through an eddy viscosity \citep{PhysRevFluids.9.094606, hamba1995analysis, hamba2004nonlocal}.
In principle, the interaction between the turbulent fluctuations $\boldsymbol{U}'$
and the perturbation field $\boldsymbol{u}$ may involve a nonlocal dependence in time,
so that the effective turbulent stress would contain a memory kernel that depends on
the past history of the perturbation.
In the present work, we neglect such memory effects and adopt a steady eddy-viscosity
closure that depends only on the instantaneous mean strain.

The herein adopted eddy-viscosity model is an approximation to the true Reynolds stress dynamics, and linearization about this model introduces an additional simplification. Nonetheless, this approach has been successful in capturing the linear dynamics, including transient growth, and in explaining the organization and scaling of coherent structures in turbulent flows \citep{del2006linear, cossu2009optimal, vadarevu2019coherent}. 
For the present purposes, we model the eddy viscosity using the analytic expression by \citet{cess1958survey},

\vspace{-6pt}
\begin{equation}
\label{eqn:cess}
    \frac{\nu_t(y)}{\nu}= \frac{1}{2}\left\{1+\frac{\kappa^2 \operatorname{Re}_\tau^2}{9}\left(2y-y^2\right)^2\left(3-4y+2y^2\right)^2 \Big[1-\exp \left[(|1-y|-1) \operatorname{Re}_\tau / A\right]\Big]\right\}^{1 / 2}+\frac{1}{2}.
\end{equation}
The von K\'arm\'an constant is $\kappa=0.426$ and $A=25.4$.
The computation of the turbulent mean velocity profile $\mU(y)$ involves integration of $(y-1)\mU_{\tau} Re_{\tau}/(\nu_t/\nu)$ from $0$ to $y$.

\subsection{The Reynolds-averaged adjoint equations and the mean DOD}

The solution of the averaged linearized forward equations (\ref{Eqn:LRANS}) from the previous section can be written as the linear map, 

\vspace{-6pt}
\begin{equation}
    \ea{\boldsymbol{u}}_{t_m} = \mathcal{A}\ea{\boldsymbol{u}}_0,
\end{equation}
where $\mathcal{A}$ is a linear operator acting on the space of divergence-free vector fields with sufficient smoothness, representing the evolution of the perturbation from the initial condition $\ea{\boldsymbol{u}}_0$ to the measurement time $t_m$.
The adjoint evolution $\mathcal{A}\ad$ is then derived from the forward-adjoint duality relation,

\vspace{-6pt}
\begin{equation}
    \ip{ \mathcal{A}\ea{\boldsymbol{u}}_0}{ \ea{\boldsymbol{u}}\ad_{t_m} } = \ip{ \ea{\boldsymbol{u}}_0}{\mathcal{A}\ad\ea{\boldsymbol{u}}\ad_{t_m} },
\end{equation}
where the inner product is defined as $\ip{\boldsymbol{f}}{ \boldsymbol{g}} := \int_V \boldsymbol{f}^H\boldsymbol{g} dV$ and $\ea{\boldsymbol{u}}\ad$ is the adjoint of the ensemble-averaged velocity perturbation.
The Reynolds-averaged adjoint Navier-Stokes (RA-ANS) equations for the modelled mean perturbation fields $\mea{\boldsymbol{u}}$, are derived from this duality relation using integration by parts, and are given by,

\vspace{-6pt}

\vspace{-6pt}
\begin{subequations}
    \label{Eqn:ANS}
    \begin{eqnarray}
    &\partial_{\zeta} \mea{\boldsymbol{u}}\ad - \mU \partial_x \mea{\boldsymbol{u}}\ad + d_y\mU \mea{u}\ad \boldsymbol{e}_y  = \boldsymbol{\nabla} \mea{p}\ad + \nu_t \nabla^2 \mea{\boldsymbol{u}}\ad + \left(d_y\nu_t\right) \partial_y \mea{\boldsymbol{u}}\ad - \left(d_y^2\nu_t\right) \mea{v}\ad \boldsymbol{e}_y, \qquad \\
    & \boldsymbol{\nabla} \cdot \mea{\boldsymbol{u}}\ad = \boldsymbol{0}, \qquad 
    \mea{\boldsymbol{u}}\ad_{t_m} = \boldsymbol{\phi}\ad.
    \end{eqnarray}
\end{subequations}
The first equation is thus the eddy-viscosity modelled version of the ensemble-averaged adjoint momentum equation,

\vspace{-6pt}
\begin{equation}
    \partial_{\zeta} \ea{\boldsymbol{u}}\ad - \ea{\boldsymbol{U}} \cdot \nabla \ea{\boldsymbol{u}}\ad + \nabla \ea{\boldsymbol{U}} \cdot \ea{\boldsymbol{u}}\ad = \boldsymbol{\nabla} \ea{p}\ad + \nu \nabla^2 \ea{\boldsymbol{u}}\ad + \ea{\boldsymbol{U}^{\prime}\cdot\nabla\boldsymbol{u}\ad - \nabla\boldsymbol{U}^{\prime}\cdot\boldsymbol{u}\ad}.
\end{equation}
The adjoint equations are solved in backward time, $\zeta = t_m-t$. The appearance of the transposed advection operator $\nabla \ea{\boldsymbol{U}} \cdot \ea{\boldsymbol{u}}\ad$ fundamentally distinguishes the adjoint dynamics from the forward system \citep{wang_wang_zaki_2022}. 
In addition, the correlation term $\ea{\boldsymbol{U}^{\prime}\cdot\nabla\boldsymbol{u}\ad - \nabla\boldsymbol{U}^{\prime}\cdot\boldsymbol{u}\ad}$ represents the interaction between forward fluctuations and the adjoint field. 
Such forwardâ€“adjoint correlations are rarely considered in turbulence modeling, and their statistical closure remains largely unexplored.
A comparison of equations~(\ref{Eqn:LRANS}) and~(\ref{Eqn:ANS}) reveals that the eddy-viscosity closure is \emph{not} self-adjoint in the presence of mean shear. The forward eddy-viscosity term in~(\ref{Eqn:LRANS}) is $\nu_t\nabla^2\mea{\boldsymbol{u}} + \left(d_y\nu_t\right)(\partial_y\mea{\boldsymbol{u}}+\nabla\mea{v})$, whereas the corresponding term in the adjoint~(\ref{Eqn:ANS}) is $\nu_t\nabla^2\mea{\boldsymbol{u}}\ad + \left(d_y\nu_t\right)\,\partial_y\mea{\boldsymbol{u}}\ad - \left(d_y^2\nu_t\right)\,\mea{v}\ad\boldsymbol{e}_y$. The extra term $-\left(d_y^2\nu_t\right)\,\mea{v}\ad\boldsymbol{e}_y$ arises from integration by parts of the wall-normal gradient contribution and reflects the wall-normal inhomogeneity of $\nu_t(y)$. In the limiting case of a spatially uniform eddy viscosity ($d_y\nu_t = 0$), both operators reduce to $\nu_t\nabla^2$, which is self-adjoint; it is the wall-normal variation of $\nu_t$ driven by mean shear that breaks the self-adjoint structure of the closure.

\begin{figure}
    \centering    \includegraphics[width=0.8\linewidth]{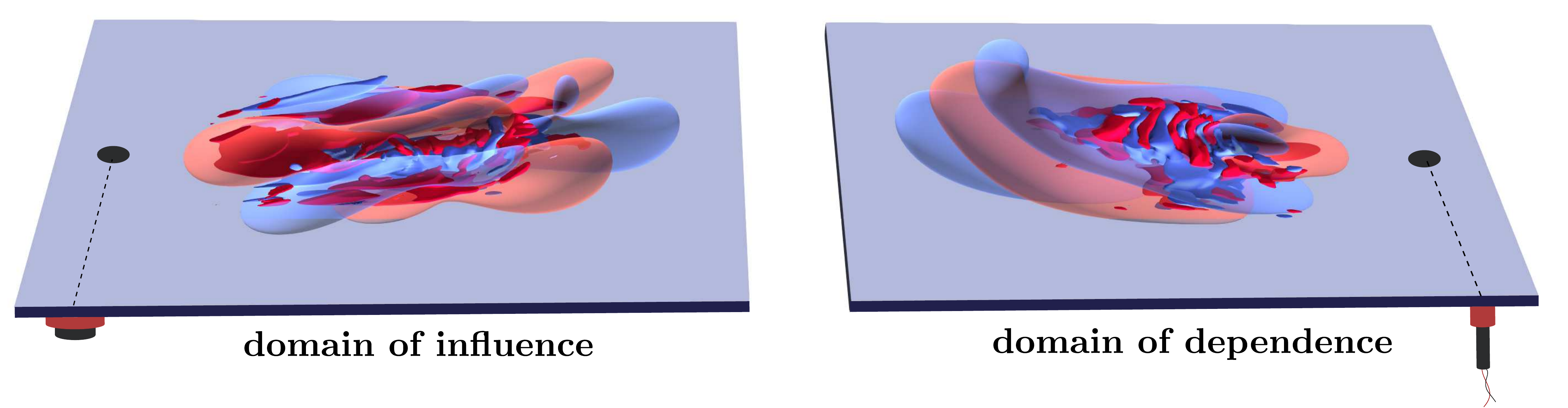}
    \caption{Examples of the instantaneous and mean domains of influence (DOI) from the linearized Navier-Stokes and domain of dependence (DOD)  from the adjoint. Opaque isosurfaces: DOI and DOD computed from one realization of channel flow turbulence. Transparent isosurfaces: mean DOI and DOD.}
    \label{fig:schematic}
\end{figure}

To demonstrate the physical meaning of the adjoint field, we prescribe its terminal condition $\boldsymbol{u}\ad_{t_m} = \boldsymbol{\phi}\ad$ as the measurement kernel, i.e.\,$\ip{\ea{\boldsymbol{u}}_{t_m}}{\boldsymbol{\phi}\ad } = \ea{m}$, and substitute into the duality relation. We obtain,

\vspace{-6pt}
\begin{equation}
    \ip{\ea{\boldsymbol{u}}_{t_m}} {\boldsymbol{\phi}\ad }= \ip{\mathcal{A}\ea{\boldsymbol{u}}_0}{\boldsymbol{\phi}\ad } = \ip{\ea{\boldsymbol{u}}_0}{\mathcal{A}\ad\boldsymbol{\phi}\ad } = \ip{\ea{\boldsymbol{u}}_0}{\ea{\boldsymbol{u}}_0\ad }.
\end{equation}
As such, the mean variation in the measurement, $\ea{m}$, is proportional to the mean initial perturbation $\ea{\boldsymbol{u}}_0$, and the proportionality is given by the mean adjoint field $\ea{\boldsymbol{u}}_0\ad$. 
The mean adjoint can therefore be interpreted as the mean DOD for the measurement kernel $\boldsymbol{\phi}\ad$. This idea is demonstrated schematically in figure \ref{fig:schematic}, where the DOI and DOD using one realization of channel flow turbulence are plotted, along with their ensemble averages, which are shown in transparent iso-surfaces.

\section{Results}
\label{sec:results}

\subsection{The mean domain of dependence}
\label{sec:influence_dependence}

We solve both the Reynolds-averaged linearized and adjoint Navierâ€“Stokes equations, RA-LNS (\ref{Eqn:LRANS}) and RA-ANS (\ref{Eqn:ANS}) respectively,
initialized using wall-stress data.  The forward equations provide the mean domain of influence (DOI) of a wall-stress perturbation, and the adjoint equations provide the mean domain of dependence (DOD) of a wall-stress measurement kernel in backward time. 
The base state for both the RA-LNS and RA-ANS equations is the mean base-flow profile $\mU(y)$, computed using the Cess eddy-viscosity model~(\ref{eqn:cess}). This profile satisfies the steady Reynolds-averaged Navier-Stokes equations for fully developed and statistically stationary channel flow, and not the instantaneous Navier-Stokes equations~(\ref{Eqn:NS}).
The initial wall-stress kernel at the wall is the derivative of a narrow Gaussian that represents $\partial u/\partial y\big|_{y=0}$.  Specifically, we use $
    \phi(\boldsymbol{x}) = \left({y^+}/{\sigma^5 \left( 2 \pi \right)^{3/2}}\right) \exp\left(-{|\boldsymbol{x}|^2}/{2 \sigma^2}\right)$,
where $\sigma$ is defined as the width of the Gaussian in viscous units.
A projection operator that enforces the solenoidal and boundary conditions is applied to the initial condition.
The forward solver is validated against the results from \citet{pujals2009note}, including optimal transient growth curves.  The adjoint solver is validated using the forwardâ€“adjoint duality relation described in \S\ref{Sec:forumulation}.
The friction Reynolds numbers that were examined are $Re_{\tau} = \{180,590,10^3,10^4\}$.
For a given friction Reynolds number, the linearized and adjoint fields are evaluated for streamwise wavenumbers $k_x^+ \in [0,0.4775]$ sampled at intervals of $\Delta k^+_x=2.5\times 10^{-3}$, and spanwise wavenumbers $k_z^+ \in [-0.5,0.5]$ sampled at intervals of $\Delta k_z^+ = 4\times 10^{-3}$.

\begin{figure}
    \centering
    \includegraphics[width = \textwidth]{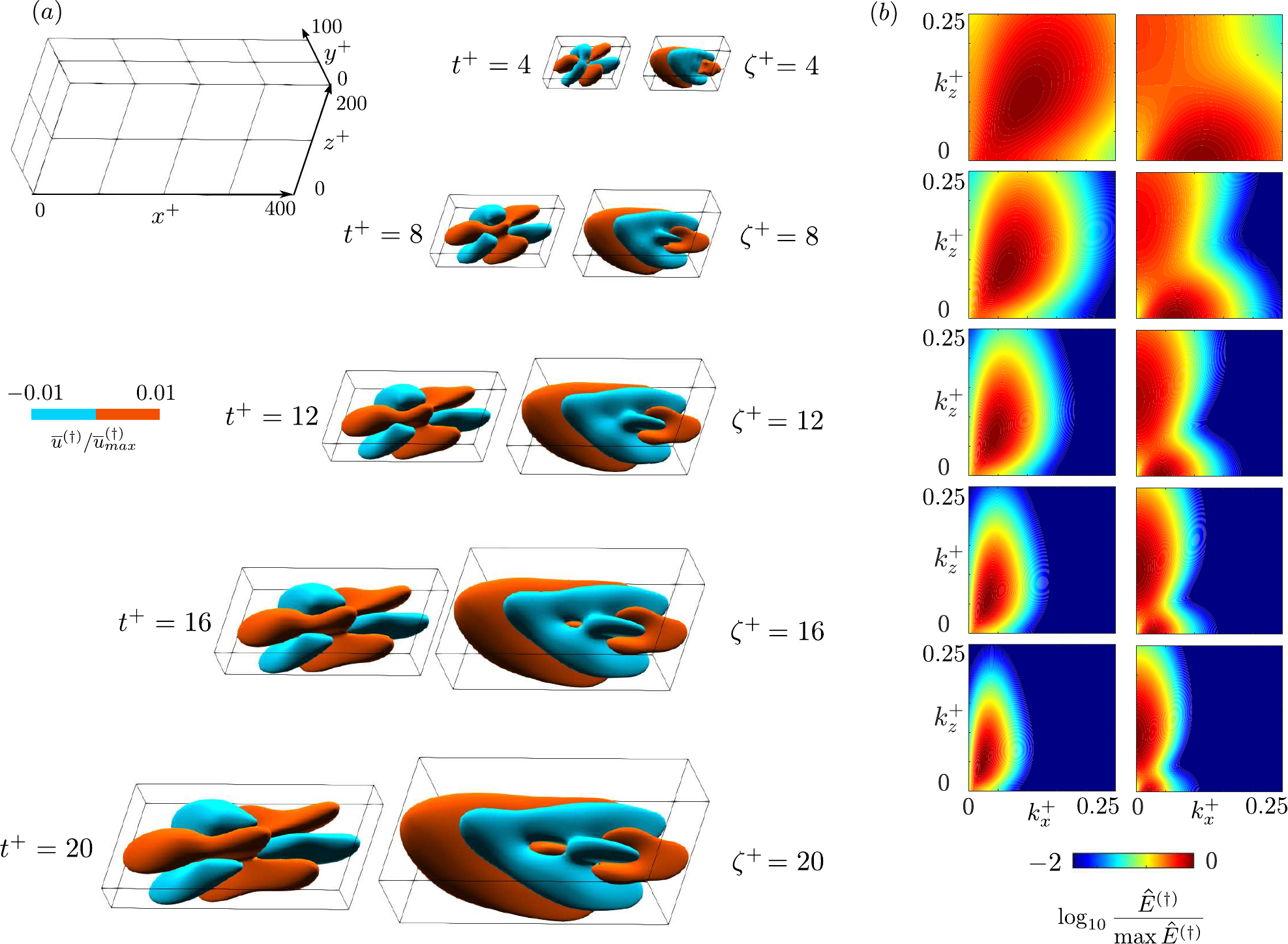}
    \caption{Comparison of the mean DOI and DOD at $Re_\tau = 10^3$. (a) The DOI (left column) and the DOD (right column) starting from the measurement kernel of $\frac{\partial u}{\partial y}$ at the wall ($y=0$). Iso-surfaces of the streamwise velocity components of these structures are shown at selected forward and backward time instances. The scale in the upper-left corner applies to all panels. (b) Contours of the energy in different wave numbers $(k_x^+, k_z^+)$, and vertically integrated across the channel, are shown for the DOI (left column) and DOD (right column) at the corresponding instances.}
    \label{fig:adjoint_forward_structures}
\end{figure}

Iso-surfaces of the normalized mean DOI and DOD are shown in figure \ref{fig:adjoint_forward_structures}$(a)$, and their energy spectra are reported in panel $(b)$.  
The results are for $Re_{\tau} = 10^3$, and those for the other Reynolds numbers are visually indistinguishable when plotted in viscous units.  
The DOI and DOD expand in forward and reverse time, respectively, while largely preserving their shapes (figure \ref{fig:adjoint_forward_structures}$(a)$).  
The qualitative difference in their form in physical space reinforces that the forward influence of a perturbation of wall stress and the sensitivity of a wall-stress measurement are not mirror images of each other.
In wavenumber space (panel $(b)$), the energy distribution of the DOI is concentrated in a single, simply connected region, characterized by a dominant wave number. In contrast, the DOD energy is distributed across two distinct regions: 
One region is associated with upstream-leaning $u$-velocity sheets, which correspond to low $k_z$ in the spectra. This region is dominant at early backward times and persists at longer backward times. The second region is associated with streamwise streaky structures. These are low $k_x$ in the spectra, and become appreciable at longer backward times.
The dynamics of these two regions will be examined after a comparison of the mean DOD and the ensemble-average from adjoint realizations. 

\begin{figure}
    \centering
    \includegraphics[width=0.95\linewidth]{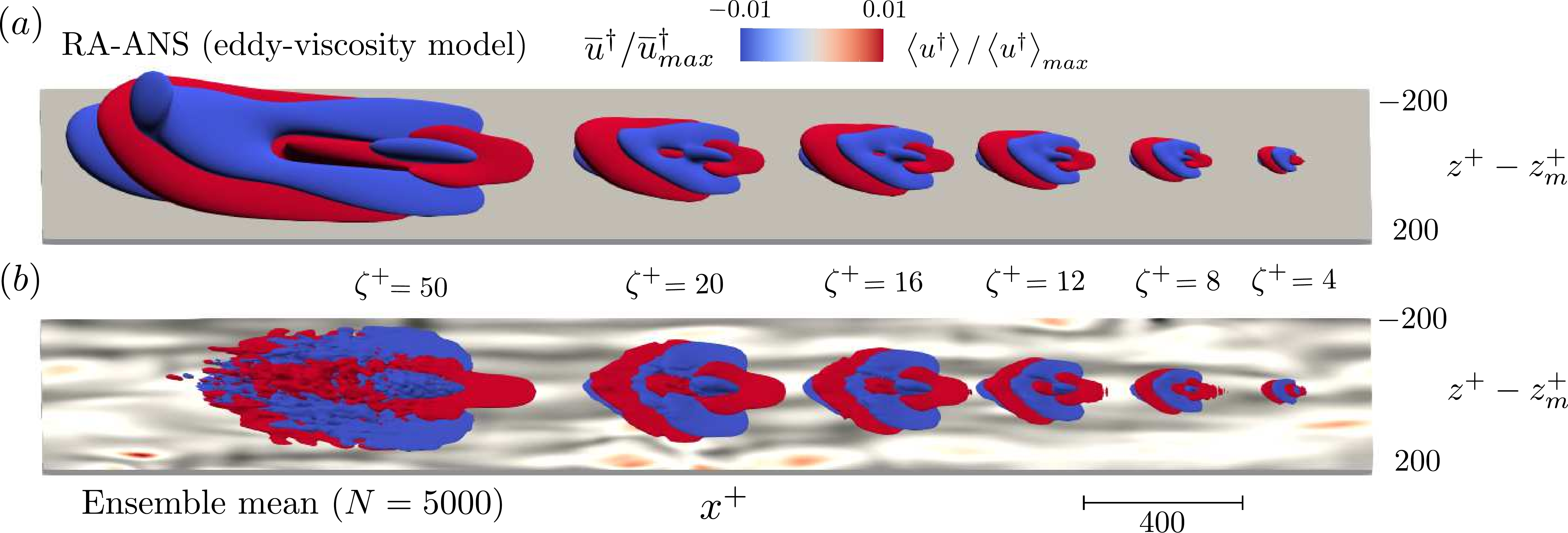}
    \caption{Comparison of the isosurfaces (a) $\mea{u}^{\dagger}/\mea{u}^{\dagger}_{max} = \pm 0.01$  and (b) $\ea{u^\dagger}/\ea{u^\dagger}_{max} = \pm 0.01$, computed from the RA-ANS and the ensemble adjoint approach (5000 samples) respectively, for backward times $\zeta^+ = \{4, 8, 12, 16, 20, 50\}$ for $Re_{\tau}=180$. The background gray color shows a sample of the underlying turbulent streamwise wall-shear stress.}
    \label{fig:compare_LST_DNS}
\end{figure}

\begin{figure}
    \centering
    \includegraphics[width=\linewidth]{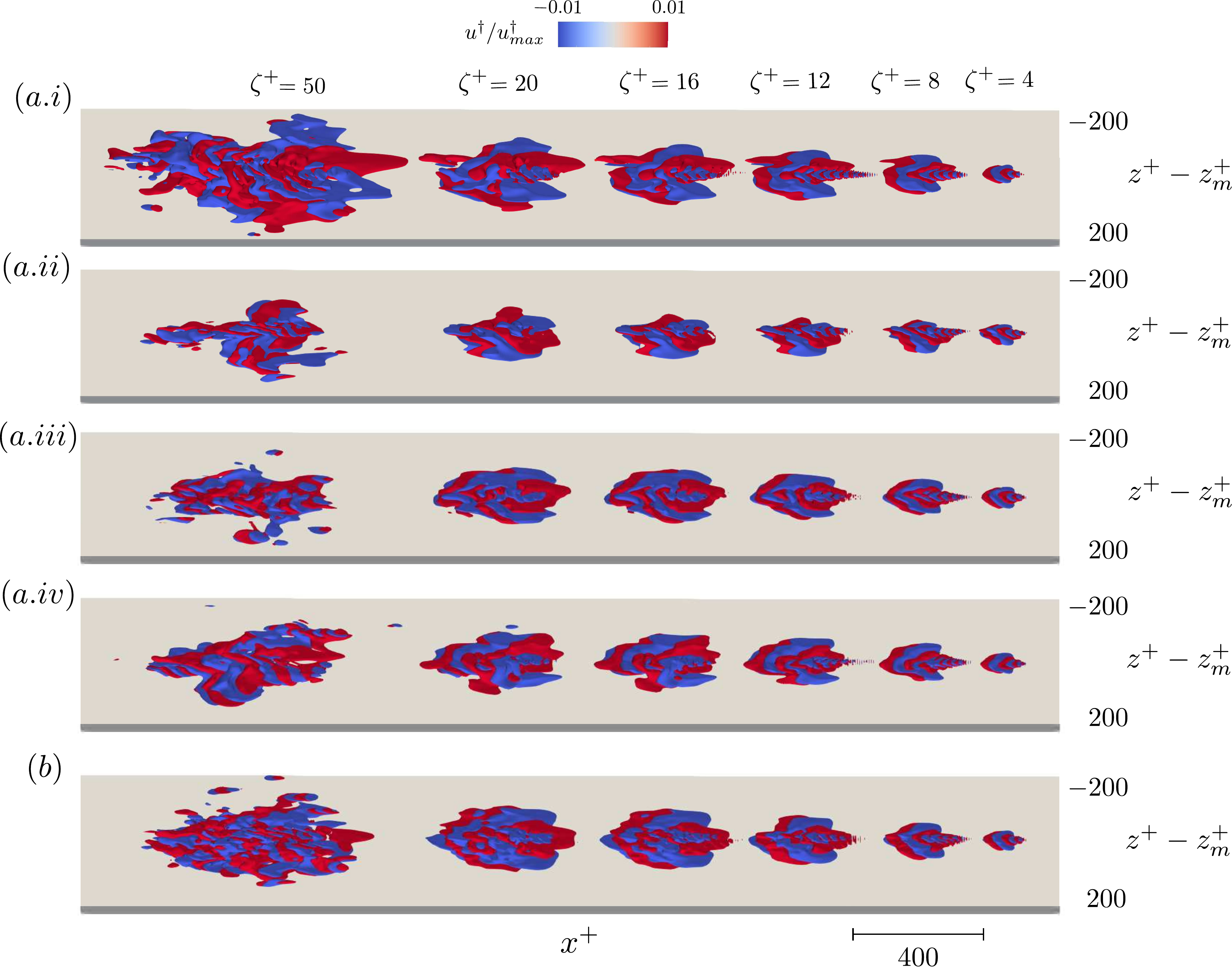}
    \caption{(a.i-iv) Instances of the domain of dependence, shown by isosurfaces of ${u}^{\dagger}/{u}^{\dagger}_{max} = \pm 0.01$, for four different realizations of the turbulent base flow, at the backward times $\zeta^+ = \{4, 8, 12, 16, 20, 50\}$ for $Re_{\tau}=180$. (b) Averaged adjoint structure using ten ensemble members.}
    \label{fig:compare_instantanesou_samples}
\end{figure}

To verify the eddy-viscosity adjoint model, we compare its prediction of the mean adjoint field to an ensemble average of 5{,}000 realizations from adjoint Navier-Stokes simulations.  
For each ensemble member, the adjoint computation adopts an independent realization of the turbulent base flow $\boldsymbol{U}\left(\boldsymbol{x},t\right)$, sampled from a long statistically stationary DNS of channel flow.  The base states were sampled at fifty instants separated by $\Delta t^+=20$, and the fields were circularly translated ten times in each of the horizontal $x$ and $z$ directions.
Figure~\ref{fig:compare_LST_DNS} shows the isosurfaces of the streamwise adjoint-velocity from both approaches. 
The structures exhibit strong visual similarity, particularly for short backward times ($\zeta^+ \le 20$). At longer times, such as $\zeta^+ = 50$, the ensemble-averaged adjoint field features high-frequency oscillations which are indicative of a lack of statistical convergence. 

In figure \ref{fig:compare_instantanesou_samples}($a$), we report four realizations of the adjoint from the ANS simulations, which should be viewed in contrast to the ensemble average in figure \ref{fig:compare_LST_DNS}.  The individual realizations exhibit appreciable variability among themselves, with complex small-scale structures due to the chaotic backward-time evolution. These small scales are, however, absent from the ensemble average in figure \ref{fig:compare_LST_DNS}, which captures the coherent mean structure. Figure \ref{fig:compare_instantanesou_samples}($b$) is the ensemble-average of ten realizations only, which should again be viewed with reference to the converged average using 5000 samples in figure \ref{fig:compare_LST_DNS}. The comparison underscores that the statistical convergence of the ensemble average becomes progressively more challenging at longer backward time.

\begin{figure}
    \centering
    \includegraphics[width=0.6\linewidth]{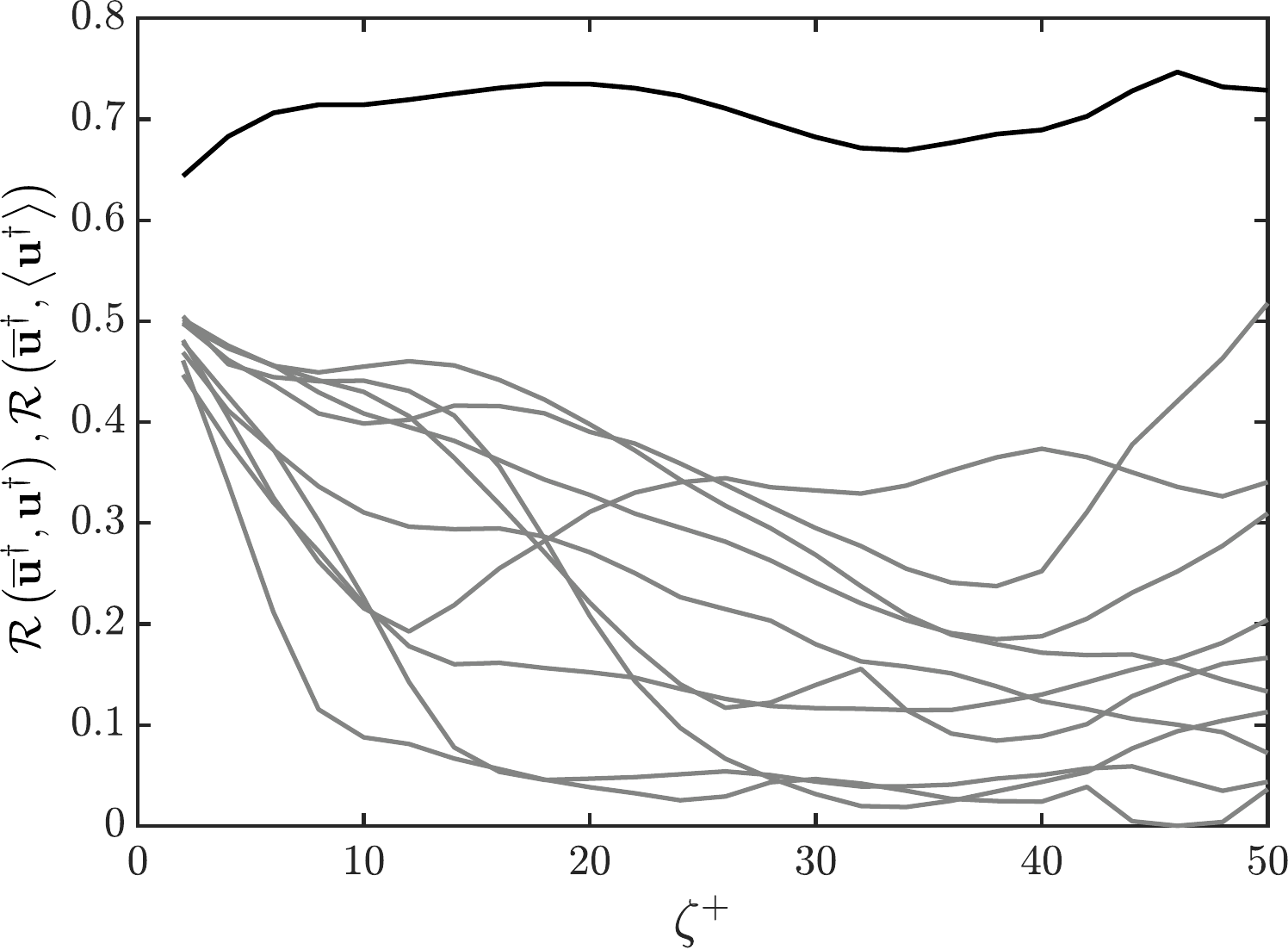}
    \caption{Spatial correlation coefficient between the RA-ANS adjoint field 
    $\bar{\boldsymbol{u}}^{\dagger}$ and the ensemble-averaged adjoint 
    $\langle \boldsymbol{u}^{\dagger} \rangle$ obtained from $5000$ realizations (black line), 
    and ten representative individual adjoint realizations 
    $\boldsymbol{u}^{\dagger}_{(n)}$ without averaging (gray lines). 
    The strong positive correlation (approximately $0.7$) with the ensemble mean, 
    contrasted with the weak correlation with individual realizations, 
    demonstrates that RA-ANS captures the ensemble-averaged sensitivity direction 
    while filtering the chaotic fluctuations of single adjoint trajectories.}
    \label{fig:correlation}
\end{figure}

In order to further quantify the alignment, or similarity, between adjoint fields, we define the spatial correlation coefficient, 
\begin{equation}
\mathcal{R}(\boldsymbol{a},\boldsymbol{b})
=\frac{\int_V \boldsymbol{a} \cdot\boldsymbol{b}\,dV}
{\left(\int_V|\boldsymbol{a}|^2\,dV\right)^{1/2}
\left(\int_V|\boldsymbol{b}|^2\,dV\right)^{1/2}}. 
\end{equation}
In figure~\ref{fig:correlation}, we report the correlation between the RA-ANS $\bar{\boldsymbol{u}}^{\dagger}$ and ten adjoint realizations $\boldsymbol{u}^{\dagger}$ (gray curves).  
These correlations are weak, often on the order of $0.1$ and at most about $0.5$, which is due to the chaotic backward-time evolution of individual adjoint realizations.  The figure also reports the correlation between the RA-ANS $\bar{\boldsymbol{u}}^{\dagger}$ and the ensemble average $\langle \boldsymbol{u}^{\dagger} \rangle$ from $5000$ realizations (black line). 
This correlation remains strongly positive, reaching approximately $0.7$ over the time interval where the ensemble statistics are converged. This result demonstrates that while individual adjoint trajectories are highly variable, the RA-ANS formulation successfully captures the dominant direction of the ensemble-averaged sensitivity.

The alignment between the RA-ANS $\bar{\boldsymbol{u}}^{\dagger}$ and the ensemble average $\langle{\boldsymbol{u}}^{\dagger}\rangle$ has implications for adjoint-variational data assimilation, where the gradient of the cost functional is furnished by the adjoint field. In that context, the adjoint field establishes the direction of steepest ascent (or descent) of the cost functional. Although individual adjoint realizations exhibit chaotic divergence and are poorly correlated with the RA-ANS field, the strong positive correlation between $\bar{\boldsymbol{u}}^{\dagger}$ and the ensemble-averaged adjoint $\langle \boldsymbol{u}^{\dagger} \rangle$ indicates that the RA-ANS solution captures the dominant gradient direction associated with the mean sensitivity. As a result, updates proportional to $\bar{\boldsymbol{u}}^{\dagger}$ are expected to produce a positive ensemble-averaged first-order variation of the measurement functional. In this sense, the RA-ANS formulation can be viewed as a regularized approximation of the ensemble sensitivity direction, analogous to a regularized gradient \citep{protas2004computational}, or a preconditioned gradient that filters out unstable fluctuations \citep{ke2026preconditioned}. While the eddy-viscosity approach does not reproduce the detailed structure of individual adjoint trajectories, it captures the large-scale sensitivity structure, which may be useful for interpreting the dependence of measurements.

\begin{figure}
    \centering
\includegraphics[width=0.8\linewidth]{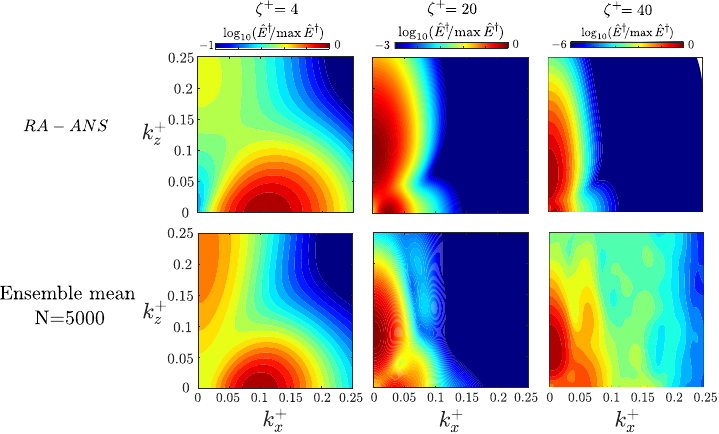}
\vspace{-1 pt}
    \caption{Energy spectra of the DOD as a function of $\boldsymbol{k}^+ = (k_x^+, k_z^+)$, integrated in the wall-normal direction for $Re_{\tau}=180$. The energy is normalized by the maximum value among all wavenumber pairs. Top: Normalized energy spectra from RA-ANS. Bottom: Normalized energy spectra from averaging $5 \times 10^3$ samples of adjoint Navier-Stokes.}
    \label{fig:fft_comparison}
\end{figure}

\begin{figure}
    \centering
\includegraphics[width=0.8\linewidth]{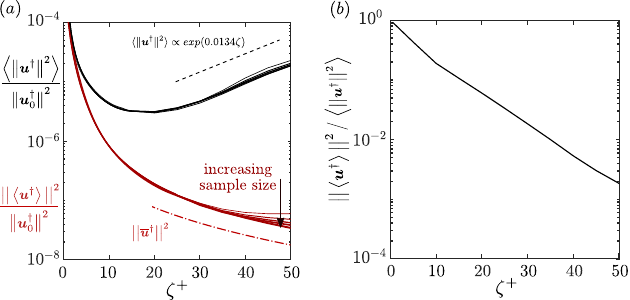}
\vspace{-1 pt}
    \caption{(a) Convergence of the ensemble adjoint method.  Squared norm of the ensemble-averaged adjoint fields ($\|\ea{\boldsymbol{u}\ad}\|^2$, red solid lines), and ensemble average of the adjoint energy ($\ea{||\boldsymbol{u}\ad||^2}$, black solid lines). Thicker lines indicate $\{1, 2, 3, 4, 5\} \times 10^3$ samples. Dashed black line marks the best exponential fit $e^{0.0134\zeta^+}$ for the averaged adjoint energy. The mean energy decay using RA-ANS is marked by the red dash-dotted line. (b) Ratio between the energy of the mean and mean energy for the adjoint field using $5 \times 10^3$ samples}.
    \label{fig:energy_evolution}
\end{figure}

A comparison in spectral space is presented in figure~\ref{fig:fft_comparison}. 
The RA-ANS prediction (top) and the ensemble-averaged adjoint spectra (bottom) exhibit similar large-scale organization, with energy concentrated in comparable regions of $(k_x^+,k_z^+)$ space.
At long times, the low wavenumbers show good agreement.
These modes correspond to the dominant large-scale structures that are most robust under ensemble averaging.
In contrast, the high-wavenumber components, which are difficult to converge in the ensemble-averaged adjoint approach, show significant discrepancies in their energy.
This discrepancy reflects the strong sensitivity of small-scale adjoint structures to chaos, which also leads to slow statistical convergence.

In figure~\ref{fig:energy_evolution}($a$), we report the norm of the ensemble-averaged adjoint field (red) and of their energy (black), plotted versus backward time.  
Convergence is shown with the number of realizations (thin-to-thick lines).  
Even though each adjoint trajectory is amplifying exponentially, the ensemble-averaged adjoint field from five thousand realizations decays through $\zeta^+ \approx 50$. 
A smaller number of samples is insufficient to yield converged statistics throughout the reported backward time horizon. 
The RA-ANS prediction of the energy $\|\mea{u}^{\dagger}\|^2$ is shown by the red dash-dotted line, and exhibits convincing agreement with the converged ensemble-averaged adjoint energy $\|\ea{\boldsymbol{u}^\dagger}\|^2$ over the time interval $20 \le \zeta^+ \le 50$.  This agreement indicates that the model captures the dominant large-scale sensitivity despite the chaotic growth of individual realizations.
In contrast, the ensemble-averaged energy $\langle\|{\boldsymbol{u}^\dagger}\|^2\rangle$ grows exponentially with time, at twice the Lyapunov exponent.  In panel ($b$), we plot the ratio of the two quantities, 
$\|\ea{\boldsymbol{u}\ad}\|^2 / \ea{||\boldsymbol{u}\ad||^2}$, 
which shows an exponential decay, thus indicating that convergence of the ensemble average becomes exponentially more difficult for longer backward times.

We verified that the RA-ANS predictions are not strongly influenced by the magnitude of the eddy viscosity by performing additional computations in which $\nu_t(y)$ was uniformly scaled by factors of $0.8$ and $1.2$ relative to the nominal Cess profile. The resulting mean DOD fields exhibit no noticeable qualitative differences from the baseline case. This test supports the view that the dominant spatio-temporal organization of the mean DOD is primarily governed by the underlying mean-shear dynamics, rather than by the precise magnitude of the turbulent diffusivity.

\begin{figure}
    \centering
    \includegraphics[width=0.7\linewidth]{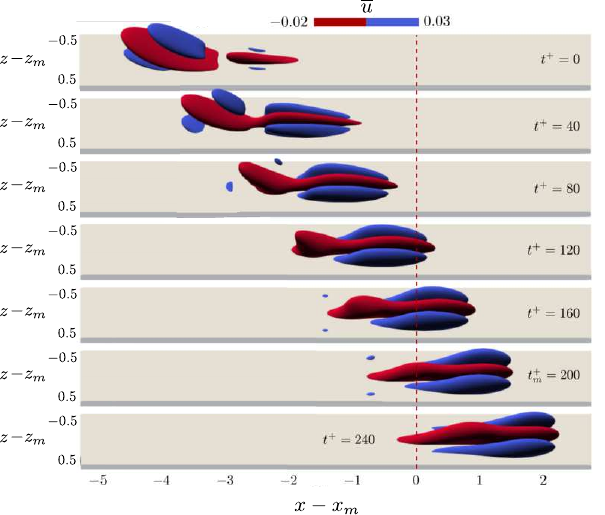}
    \caption{Forward evolution of DOD of a late-time streamwise wall shear measurement (\( t_m^+ = 200 \)) in a turbulent channel with $Re_{\tau} = 1000$, shown using isosurfaces of streamwise velocity perturbation. Measurement location is marked by a vertical dashed line.}
    \label{fig:dynamics}
\end{figure}

\begin{figure}
    \centering
    \includegraphics[width=\linewidth]{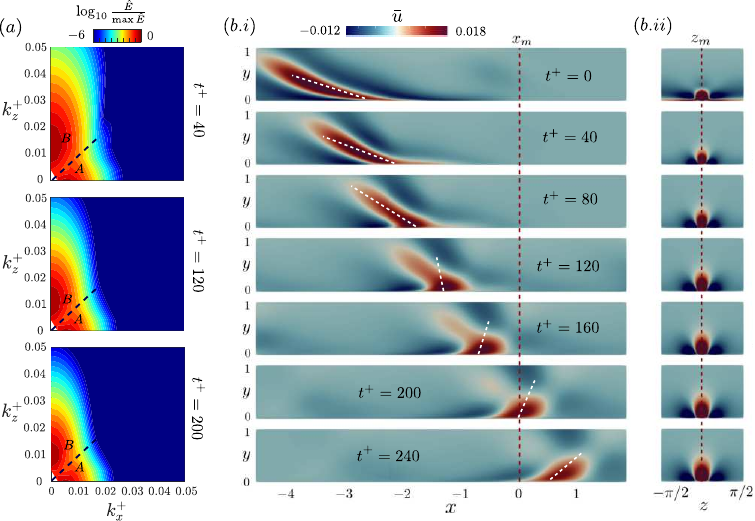}
    \caption{(a) The normalized energy spectra at three selected times $t^+=\{40,120,200\}$. Regions $A$ and $B$ denote the spectral ranges associated with the Orr-mechanism and streaky structures, respectively. (b) Forward evolution of DOD of a late-time measurement (\( t_m^+ = 200 \)). Contours of (b.i) spanwise and (b.ii) streamwise averages of $\mea{u}$, illustrating the contribution of the Orr mechanism and streaky structures to the perturbation dynamics in a turbulent channel of $Re_{\tau} = 1000$. White dashed lines in the side view mark the inclination angle of the structure.}
    \label{fig:dynamics_averaged}
\end{figure}

\subsection{Forward dynamics of the mean DOD}
To understand the dynamics of the mean DOD of a wall-stress measurement, we initialize a forward simulation with the adjoint state from a measurement at time $t_m^+ = 200$, by setting $\overline{\boldsymbol{u}}_0 = \mea{\boldsymbol{u}}^\dagger$. We then compute the forward evolution using the RA-LNS, and the results are reported in figure~\ref{fig:dynamics}.   
The DOD can be clearly separated into two components: The first is comprised of upstream-tilted structures and the second features streaky structures. 
The upstream-tilted structures evolve rapidly under the influence of the mean shear, and the dependence of the wall-stress measurement on Orr amplification in the earlier flow. The corresponding wavenumbers in the energy spectra are identified as region $A$ in figure~\ref{fig:dynamics_averaged}($a$). The peak streamwise wavenumber $k_x^+$ remains nearly constant over time, consistent with localized non-modal amplification processes. On the other hand, the streaky structures have small spanwise wavenumber $k_z^+$, and are identified in the spectra (figure ~\ref{fig:dynamics_averaged}($a$)) as region $B$. The peak $k_z^+$ becomes smaller throughout the forward evolution, indicating an elongation of the streaks. Such structures align with the most energetic modes of the forward dynamics.  Together, these two parts combine in an optimal manner to influence the measurement data at the measurement time.

To further distinguish these two components, we isolate each of them by averaging the evolved fields in the spanwise \ref{fig:dynamics_averaged}($b.i$) and streamwise \ref{fig:dynamics_averaged}($b.ii$) directions. This averaging effectively spectrally filters the field along the $k_z^+=0$ and $k_x^+=0$ axes, respectively. These two cases highlight the dominant contributions associated with the Orr-type structures and streaky motions.
The spanwise averaged $\left[ \mea{u} \right]_z$ in figure \ref{fig:dynamics_averaged}$(b.i)$ exhibits initial alignment against the mean shear and gradually rotates towards vertical alignment as time evolves, which is consistent with the Orr mechanism. The streamwise averaged $\left[ \mea{u} \right]_x$ in figure \ref{fig:dynamics_averaged}$(b.ii)$ shows a self-similar expansion while maintaining its overall shape, characteristic of streak growth dynamics.

\begin{figure}
    \centering    \includegraphics[width=\linewidth]{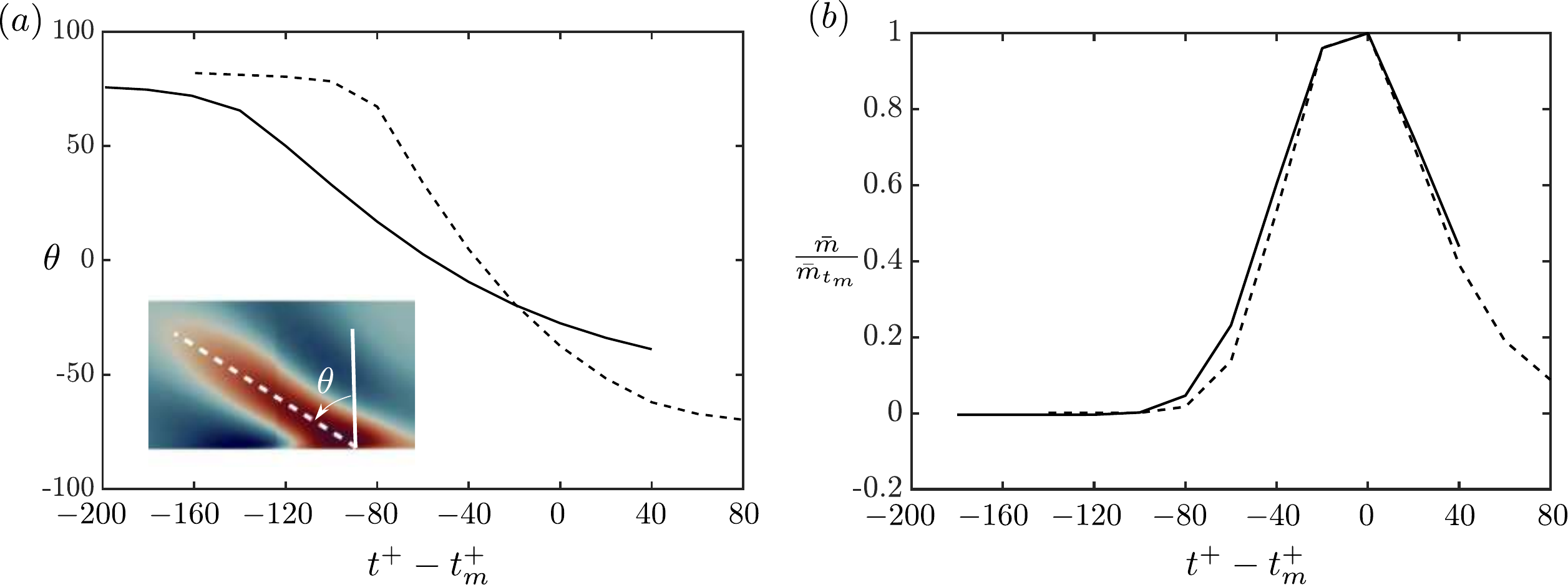}
    \caption{(a) Temporal evolution of the near-wall orientation angle $\theta$, computed from the local phase gradient of $[\bar{u}]_z$. (b) Wall shear stress at the sensor location, normalized by its value at $t_m$, from forward evolution of the DOD associated with (solid) $t_m^+ = 200$ and (dashed) $t_m^+ = 160$.}
    \label{fig:measurement_angle}
\end{figure}

For the Orr amplification, we compute the dominant orientation of the flow field from the phase of the streamwise velocity in Fourier space. Specifically, we consider the complex-valued mode shape $\hat{\mea{u}}(y,t)$ associated with the wavenumber $K^+_x$ corresponding to the maximum energy along the line $k^+_z=0$. The local phase is then defined as $\phi(y,t) = \arg\big(\hat{\mea{u}}(y,t)\big)$.
The wall-normal wavenumber is obtained from the phase gradient as $K^+_y = \partial \phi / \partial y$, which together with $K^+_x$ defines the local wavevector $\boldsymbol{K}^+ = (K^+_x, K^+_y)$. The inclination angle is defined as
\begin{equation}
    \theta =\frac{ \int_0^{y_c}\arctan\left(\frac{K^+_y}{K^+_x}\right)||\hat{\mea{u}}(y,t)||^2 dy}{\int_0^{y_c} ||\hat{\mea{u}}(y,t)||^2 dy},
\end{equation}
using $||\hat{\mea{u}}||^2$ as weights to suppress contributions from low-amplitude regions, with integration from the wall to the cut-off range $y_c = 0.2$. This procedure yields a time-resolved estimate of the dominant orientation of $\left[ \mea{u} \right]_z$ in the $(x,y)$ plane.
As shown in figure \ref{fig:measurement_angle}$(a)$, the extracted orientation angles from forward evolution of DOD from two different measurement times exhibit similar temporal dynamics: The structure is initially upstream-tilted relative to the mean shear, then progressively rotates toward a near-vertical alignment at the measurement time. 
The time dependence of the wall-shear-stress at the measurement locations, $\mea{m} = \nu\left({\partial \mea{u}(t)}/{\partial y}\right)\Big\vert_{\boldsymbol{x}_m}$, are plotted in figure~\ref{fig:measurement_angle}(b) normalized by the values at the measurement times, $t_m^+ = 200$ and $t_m^+ = 160$. The curves reflect the transient amplification of the fields as they approach the measurement times, where the peak stress is recorded. The earlier-in-time amplification takes place as the adjoint structure undergoes reorientation by the mean shear, consistent with the Orr mechanism.  
These observations are qualitatively similar to the findings by \citet{encinar2020momentum}, where conditionally averaged ejection and sweep events display a rotation from upstream-tilt to near-vertical configurations.  Note, however, that the adjoint DOD computes the optimal precursor structure to maximize the measurement data  \citep{wang2025domain}, here the wall-shear stress and not the perturbation kinetic energy. As such, there is no \emph{a priori} justification for an agreement.  
For example, unlike energy-based optimal growth, the peak amplification occurs near $\theta \approx -30^\circ$, shortly after the structure passes through vertical alignment. The difference is amply justified by the difference in the objectives, and arguably more important is the persistent importance of Orr amplification.

The two-component decomposition of the mean adjoint field is effectively a distinction of two types of information contained in wall shear-stress measurements, each associated with a different physical mechanism. 
Together, Orr amplification and lift-up define the flow regions and mechanisms, or information, that are observable from the measurements in an averaged sense, which are typically obscure in the chaotic dynamics of individual realizations of the adjoint fields.

\section{Conclusion}
\label{sec:conclusion}

Estimating the domain of dependence (DOD) of a measurement in turbulent channel flow is challenging due to the exponential growth of adjoint fields in backward time, which reflects the inherent difficulty of data assimilation in turbulence. While ensemble averaging of adjoint trajectories can, in theory, yield the mean DOD, this approach suffers from prohibitively slow convergence and overwhelming computational cost. 
Specifically, while the ensemble-averaged adjoint field should decay, the energy of individual adjoint realizations amplifies at twice the Lyapunov exponent and, as such, a very large number of adjoint trajectories are required in the ensemble averaging for convergence.  
We overcome these obstacles by introducing a linear eddy-viscosity closure to approximate the forwardâ€“adjoint correlation in the Reynolds-averaged adjoint Navierâ€“Stokes equations (RA-ANS), thus enabling a direct and efficient computation of the mean DOD. This model bypasses the exponential divergence of individual adjoint trajectories, and its accuracy was demonstrated against the expensive ensemble average sensitivity up to a backward time $\zeta^+=\mathcal{O}(40)$.  

The present work establishes a connection between several concepts that are typically considered separately: adjoint-based sensitivity analysis, the Lyapunov-driven divergence of chaotic trajectories, and the non-normal transient growth mechanisms that organize turbulent structures. In particular, the results highlight how the loss of information associated with long adjoint time horizons coexists with a coherent ensemble-averaged sensitivity that can be captured through appropriate regularization using eddy viscosity for the ensemble-averaged adjoint equations.

Our results reveal that the mean DOD exhibits a universal structure across a wide range of Reynolds numbers, which comprises two distinct dynamical components: an upstream-tilted part associated with the Orr mechanism, which is rapidly reoriented by the mean shear to influence wall observations, and streamwise-elongated streaks that reflect optimal energy growth over longer timescales. These components combine in a dynamically optimal way to shape the DOD.  

The DOD displays a marked asymmetry relative to its forward counterpart, the domain of influence (DOI), reflecting the non-self-adjoint nature of the governing operator in the presence of Reynolds stresses. This asymmetry highlights the limitations of interpreting observations based solely on forward dynamics. In contrast, the mean DOD provides a more appropriate framework for identifying the flow structures that influence a given measurement, as directly quantifies the backward-time sensitivity to earlier flow states.
Furthermore, the spatio-temporal and spectral characteristics of the DOD are important because they describe the information content in wall-stress measurement, which can be extracted using data assimilation.

\par\bigskip\noindent
\textbf{Funding.} The authors acknowledge financial support from the National Science Foundation (grant 2332057) and the Office of Naval Research (grant N00014-25-1-2290).

\par\medskip\noindent
\textbf{Declaration of interests.} 
The authors report no conflict of interest.
 
\par\medskip\noindent
\textbf{Author ORCIDs.} \\
Qi Wang  \url{https://orcid.org/0000-0001-7393-9922}\\
Tamer A. Zaki  \url{https://orcid.org/0000-0002-1979-7748}

\bibliographystyle{jfm}
\bibliography{Manuscript}

\end{document}